\documentclass[preprint,usenatbib,usegraphicx]{mn2e}
\usepackage{longtable}

\begin{document}

\title
[NGC 1407 and NGC 1400 globular clusters]
{An imaging study of the globular cluster systems of NGC 1407 and NGC 1400}
\author[Forbes et al.]
{Duncan A. Forbes$^{1}
\thanks{dforbes@astro.swin.edu.au}$,
Patricia S\'anchez-Bl\'azquez$^{1}$,
Anna T. T. Phan$^{1}$, 
\newauthor
Jean P. Brodie$^{2}$, Jay Strader$^{2}$, Lee Spitler$^{1,2}$\\
$^{1}$Centre for Astrophysics \& Supercomputing, Swinburne University,
Hawthorn, VIC 3122, Australia\\
$^{2}$Lick Observatory, University of California, CA 95064, USA}

\maketitle

\begin{abstract}

We present wide-field Keck telescope imaging of the globular cluster
(GC) systems around NGC 1407 and NGC 1400 in the Eridanus galaxy cloud. This is
complemented by Hubble Space Telescope images from the Advanced Camera for
Surveys of NGC 1407 and Wide Field and Planetary Camera 2 images of NGC
1400. We clearly detect bimodality in the GC colour distribution of NGC
1407. The blue GC subpopulation has a mean
colour of $B-I$ = 1.61 and a relative contribution of around 40\%,
whereas the red subpopulation with $B-I$ = 2.06 contributes 60\% to the
overall GC system. Assuming old ages, this corresponds to [Fe/H]
= --1.45 and --0.19. Both subpopulations are intrinsically broad
in colour (indicating a range in ages and/or metallicities), with the red 
subpopulation being broader than the blue. 
The GC colour distribution for NGC 1400 is less clear cut than for NGC
1407, however, we also find evidence for a bimodal
distribution. 
We find the NGC 1407 red GCs to be 20\% smaller in size than
the blue ones. This is  consistent with the expectations of mass
segregation in an old coeval GC system. Half a dozen 
large objects (20--40 pc), with GC-like colours are identified,
which are probably background galaxies.

The HST
data sets allow us to probe to small galactocentric radii. Here we find
both GC systems to possess a GC surface density distribution which is
largely constant in these inner galaxy regions. We fit isothermal-like
profiles and derive GC system core radii of 9.4 kpc for NGC 1407 and 5.8
kpc for NGC 1400. For NGC 1407 we are able to separate the surface density
distribution into blue and red subpopulations, giving 17.8 and 7.6 kpc
respectively. Outside this central region, the radial profile of the
GC surface density is similar to that of the galaxy light for NGC 1407 but
it is flatter for NGC 1400. The mean GC magnitude appears to be
constant with galactocentric radius. 
We find that for both galaxies, the GC systems have a
similar ellipticity and azimuthal distribution as the underlying galaxy
starlight.
A fit to the GC luminosity function gives a distance
modulus of 31.6, which is in good agreement with distances based
on the Faber-Jackson relation and the Virgo
infall corrected velocity. This distance lies at the midpoint of recent surface
brightness fluctuation distance measurements. 

\end{abstract}

\section{Introduction}

Early-type galaxies can host hundreds to many
thousands of globular clusters (GCs). The properties of 
GC systems provide important clues about the
formation and evolutionary history of their host galaxy. This
motivation has prompted numerous imaging studies of early-type 
galaxy GC systems over the years.   
Imaging of GC systems began with wide-area photographic plates
which defined the global structural properties of GC systems (see
Harris \& Racine 1979). 
CCD imaging lead to improvements in photometry
and the first detections of colour bimodality. 
However, CCDs suffered from a small field-of-view so
studies were often limited to the central regions, in addition
contamination from background galaxies and foreground stars was a
significant issue under 1 arcsec seeing conditions Harris (1991).
The introduction of the Wide Field and Planetary Camera 2 (WFPC2)
camera on the Hubble Space
Telescope had a large impact on GC studies. Its high spatial
resolution meant that it effectively identified GCs out to
Virgo-like distances, hence reducing contamination levels to near
zero. Accurate photometry led to colour bimodality detections in the
GC systems of many/most early-type galaxies (e.g. \citealt{b:forbes}; 
Gebhardt \& Kissler-Patig 1999; Larsen et al. 2001; Kundu \& Whitmore 2001). 
The Advanced  Camera for
Surveys (ACS) camera with its increased 
area and sensitivity over WFPC2 is some ten times more effective 
for GC studies. 
Here we present some of the first results on GC systems using the
 ACS on the Hubble Space Telescope. This is complemented
with some older WFPC2 and wide-field Keck
telescope imaging. Here we study the GC systems associated with
NGC 1407 and NGC 1400.

NGC 1407 (E0) is the brightest group galaxy (BGG) in an X-ray bright 
group (Osmond \& Ponman 2004; Brough et al. 2005, in prep.), which lies 
within the larger 
Eridanus cloud.  NGC
1400 (S0-) is some 10 arcmins away from NGC 1407 in the same group. 
Although the two galaxies have quite different
heliocentric velocities (V = 1783 km/s for NGC 1407 and V = 546 km/s for NGC
1400), both are thought to lie at the same distance 
\citep{b:tonry}. Based
on galaxy velocities, \citet{b:gould} speculated that NGC 1407 lies
at the centre of a massive dark halo with M/L $\sim$ 3000. 
NGC 1407 has an extinction-corrected
magnitude of M$_V$ = --21.86 and effective radius of 1.2\arcmin
~(7.34 kpc; 
\citealt{b:bender}). It lies on the
Fundamental Plane and shows no obvious signs of morphological
fine structure \citep{b:michard}, but does reveal a young central
stellar population of $\sim$2.5 Gyrs (Denicolo et al. 2005). 
NGC 1400 has a magnitude of M$_V$~=~--20.63
and no obvious fine structure, but has a fairly large residual
of +0.3 dex with respect to the Fundamental Plane \citep{b:prugniel}. 

Motivated by the velocity differences between NGC 1407
and NGC 1400, \citet{b:perrett} undertook a study of the 
GC systems of the two galaxies to measure 
the peak of the GC luminosity functions. 
They confirmed that NGC 1400 lies
at a similar distance to NGC 1407,
despite its large peculiar velocity. They also 
estimated the specific frequency $S_N$ (number of GCs per unit
starlight) of NGC 1407 to be 4.0 $\pm$ 1.3 and the total GC 
population to be 2641 $\pm$ 443. In their observations they
detected 629 $\pm$ 76 GCs using the Washington $T_1$-band 
and 556 $\pm$ 61 GCs using the Kron-Cousins I-band. 
They noted that the GC system radial profile follows the
galaxy halo light profile unusually closely. They did not detect bimodality in 
their ($T_1 - I$) colour distribution. For NGC 1400, they derived $S_N$ = 5.2
$\pm$ 2.0 based on 106 observed GCs, and again did not detect
bimodality. 

We detect bimodality in the GC colour
distribution in both galaxies and discuss the
structural properties of the GC systems. In Sections 2 and 3  we present
imaging from the ACS and WFPC2 cameras 
respectively. In Section 4 we present the wide-field Keck
imaging. The results for NGC 1407 and NGC 1400 are given in
Sections 5 and 6. Our discussion and
conclusions are given in 
Section 7. In Appendix A
we give GC candidate lists which include coordinates, magnitudes and
colours. Coordinates are only accurate to a few arcsecs in a
relative sense, less so on an absolute scale. 
In this paper we assume a distance to NGC 1407 and NGC 1400 of 
21 Mpc, where 1 arcmin corresponds to 6.12 kpc. 


\section{ACS Observations and Data Reduction}

\begin{figure}
    \resizebox{0.9\hsize}{!}{\includegraphics{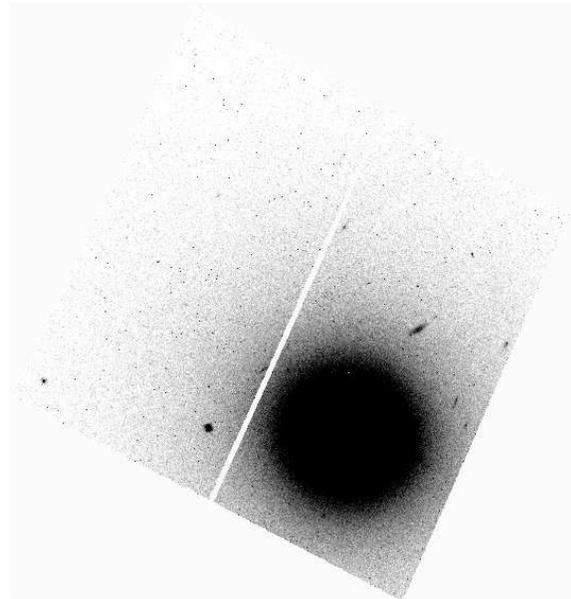}}
    \caption{HST/ACS image of NGC 1407 image in the F435W filter.
    The size of the field is 3.5\arcmin $\times$ 3.5\arcmin ~with North at the 
    top and East to the left of the frame.}  
    \label{f:acs}
\end{figure}

Advanced Camera for Surveys (ACS) images of NGC 1407 were taken
in the F435W and 
F814W filters
(proposal ID = 9427, PI = Harris, see also Harris et al. 2005).
The exposure times were 1500s in F435W and 680s in F814W. 
The galaxy centre was offset about 1\arcmin ~from the
centre of the ACS field-of-view. The images were downloaded from the HST 
archive, reprocessed using the `on-the-fly' software and then 
drizzled. Figure \ref{f:acs} shows the F435W image.

\subsection{Object Detection and Selection}

The {\sc daophotii} software package \citep{b:stetson} was used to
detect GC candidates and measure their magnitudes.  Objects with
roundness values between --1 to 1 and sharpness from 0.5 to 1.5 were
selected. We employed a detection threshold of $\sim$4, which is the
product of the {\sc find-sigma} parameter and the appropriate gain ratio.

\begin{figure}
    \resizebox{0.9\hsize}{!}{\includegraphics[angle=-90]{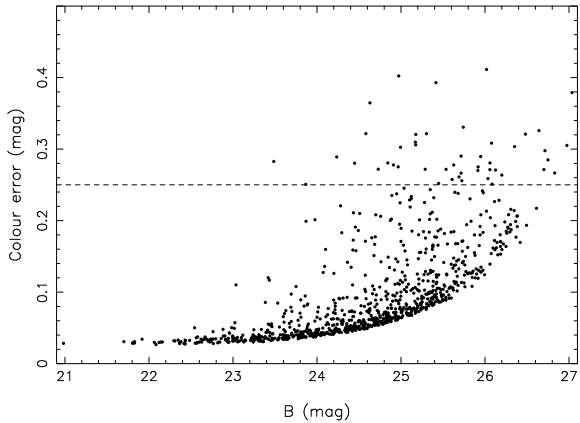}}
    \caption{The colour error as a function of $B$ magnitude for the NGC 1407
    HST/ACS data. The dashed line shows where the selection cut
of $\pm$0.25 mag. was 
    applied.}  
    \label{f:hcolourerror}
\end{figure}

Magnitudes were measured in a 5-pixel radius aperture. Aperture corrections, 
from 5 to 10 pixels, were then calculated from several bright GCs in the frame. 
Corrections from 10 pixels to the total light were taken from Sirianni et 
al. (2005). We then used the transformation equations, with colour terms, 
from Sirianni et al. (2005) to convert our 
instrumental F435W and F814W 
magnitudes into the standard Johnson B and I-band system.
Galactic extinction corrections of A$_B$ = 0.282 and A$_I$ = 0.13 were
then applied.  After applying a color selection of 1.2 $< (B-I) <$ 2.5
(which corresponds to --2.5 $<$ [Fe/H] $<$ 1.0 using the relation of
\citealt{b:barmby}), we inspected the images and visually removed
obvious contaminants (such as background galaxies).  This reduced the data
set to 952 potential globular clusters. A colour error cut of
$\pm$0.25 mag. was then applied, leaving 903 objects.  The behaviour
of the colour error for candidate GCs is shown in Figure
\ref{f:hcolourerror}.  The remaining contamination rate of foreground
stars or background galaxies is statistically insignificant given the
high spatial resolution of the ACS.  We did not detect any GCs
interior to a galactocentric radius of 1\arcsec.

\begin{figure}
    \resizebox{0.9\hsize}{!}{\includegraphics[angle=0]{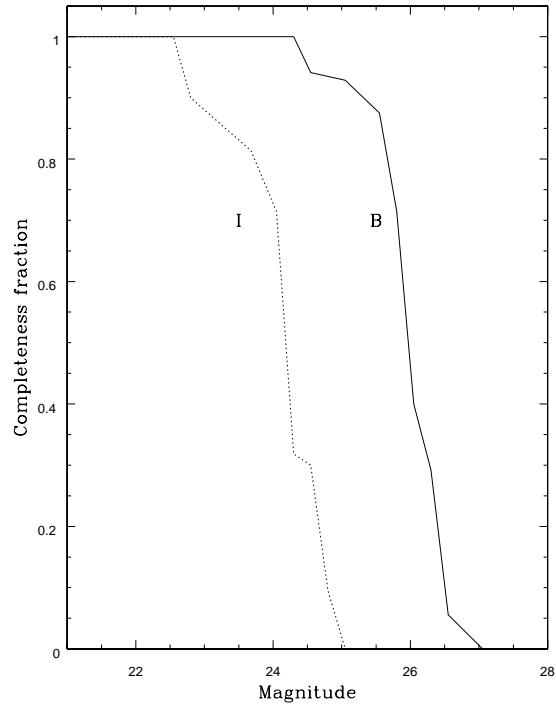}}
    \caption{GC Completeness functions. The solid line shows the
B-band completeness and the dashed line the I-band completeness
function. The 50\% completeness levels are B $\sim$ 26.0 and I
$\sim$ 24.2.  
    }  
    \label{f:complete}
\end{figure}

We have performed completeness tests by generating a large number
of artifical objects with sizes and colours that resemble
GCs. These are then randomly inserted into the F435W and
F814W ACS images using the ADDSTAR task. We maintained a density
of objects similar to that of the original data so that additional
crowding did not affect our results. 
These were then recovered using the same object
detection process as described above. Finally, the test was run three
times to improve the statistical significance. 
In Figure \ref{f:complete}
we show the recovered completeness fraction of artifical objects
as a function of B and I magnitude. The plot shows that our 50\%
completeness limits are B $\sim$ 26.0 and I $\sim$ 24.2. 

\begin{figure}
    \resizebox{0.9\hsize}{!}{\includegraphics[angle=-90]{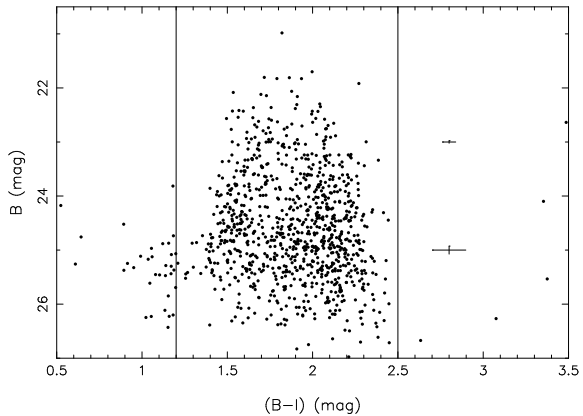}}
    \caption{Colour-magnitude diagram for NGC 1407 HST/ACS globular 
    clusters. Our colour selection is indicated by vertical lines. Typical magnitude and colour errors are shown. Bimodality at $B-I \sim$ 1.6 and 2.0 can be seen.}  
    \label{f:hcolour}
\end{figure}

A colour-magnitude diagram of the ACS GCs is shown in
Figure \ref{f:hcolour}. It shows evidence for bimodal colours, with the
brightest GCs having $B \sim$ 22. As can be seen in the figure, there is one
object which is significantly brighter than all the others, with $B$ = 20.98. 
Although $\sim$ 1 mag. brighter than the other selected objects,
A spectrum from the Keck telescope indicates that 
it has a velocity consistent with that of NGC 1407 (Cenarro et
al. 2005). Its
magnitude (M$_B$ = --10.62) is similar to that of Omega Cen in
our Galaxy. Visually it appears similar to other GCs in the ACS
image, it is not obviously a dwarf galaxy. 
Its effective radius (of $\sim$ 4 pc) is consistent with a GC at the
distance of NGC 1407 and not an Ultra Compact Dwarf
\citep{b:drinkwater}. Thus
except for this interesting object, our data set has an effective bright 
magnitude limit of $B \sim$ 21.5. The faint limit is $B \sim$ 26.

Objects with colours $B-I$ $<$ 1.2 and $B-I$ $>$ 2.5 are likely to be 
faint stars or compact galaxies. The small number of such objects 
indicates that the contamination level in our selected colour 
range (1.2 $<$ $B-I$ $<$ 2.5) is likely to be quite small. 

\section{WFPC2 Observations and Data Reduction}

\begin{figure}
    \resizebox{0.9\hsize}{!}{\includegraphics{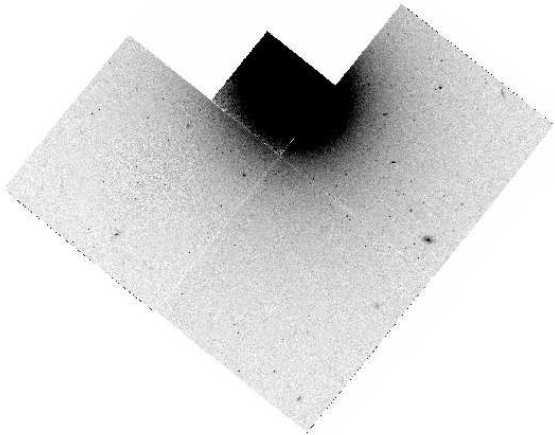}}
    \caption{HST/WFPC2 image of NGC 1400 image in the F814W filter.
    North is at the top and East to the left of the frame.}  
    \label{f:wfpc2}
\end{figure}

Wide Field and Planetary Camera 2 images of NGC 1400 were taken in the F555W and F814W filters
(proposal ID = 5999, PI = Phillips). 
The data consisted of three 160s F814W images and one 160s F555W image. 
The pipeline-reduced images were downloaded from the HST archive
and the multiple F814W images were average-combined. The four
WFPC2 CCDs were mosaiced together using the {\sc mosaic} task in
{\sc STSDAS}.
Figure \ref{f:wfpc2} shows the combined F814W image. 
To aid in the detection of globular cluster candidates on the PC image, we
fitted and subtracted the galaxy isophotes using the {\sc ellipse} task in 
{\sc iraf} allowing the centre, ellipticity and position angle to 
vary. No objects were detected within 5\arcsec ~of the galaxy
centre. 

\subsection{Object Detection and Selection}

The detection of globular clusters was carried out using the 
{\sc daophotii} package. Detection was based solely on  
the F814W image, which had a higher signal-to-noise ratio than
the F555W image but also, more importantly, was largely devoid of
cosmic rays. 
Detection criteria consisted of a roundness range of --1 to 1, a sharpness
range of 0.2 to 1 and a threshold of 4 times the background. Objects in the PC image were detected
separately from the WFC images due to the different pixel scales and background
levels. This gave over 1,000 potential GCs. 

Photometry of the detected objects was carried out for both the F555W and F814W
images using the coordinates of the objects detected in the F814W image via
the {\sc qphot} task in the {\sc apphot} package. Zeropoints and
the conversion to Johnson $V$ and $I$ bands were
taken following the method described in \citet{b:forbes} with Galactic
extinction corrections of A$_V$ = 0.21 and A$_I$ = 0.13. 

\begin{figure}
    \resizebox{0.9\hsize}{!}{\includegraphics[angle=-90]{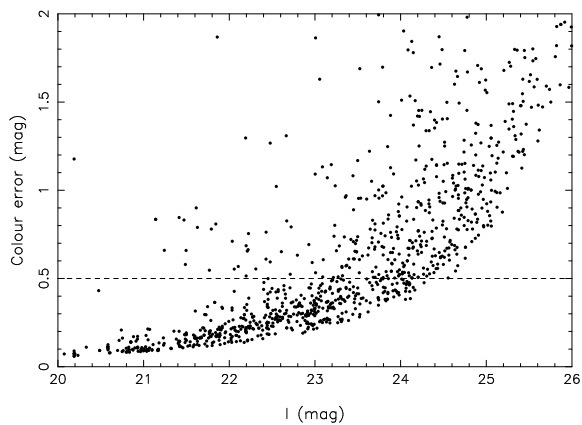}}
    \caption{The colour error as a function of $I$ magnitude for the NGC 1400
    HST/WFPC2 data. The dashed line shows where the selection cut
of $\pm$0.5 mag. was 
    applied.}  
    \label{f:wcolourerror}
\end{figure}

\begin{figure}
    \resizebox{0.9\hsize}{!}{\includegraphics[angle=-90]{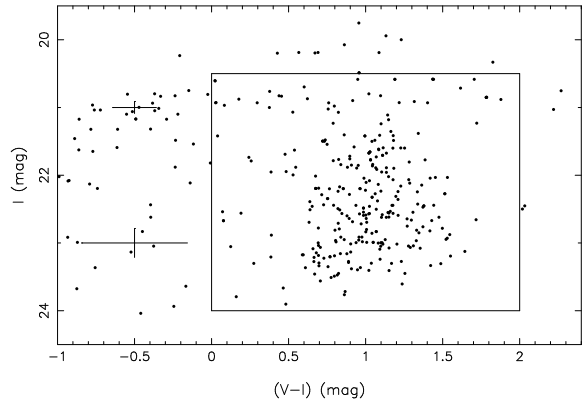}}
    \caption{Colour-magnitude diagram for objects in the NGC 1400
HST/WFPC2 imaging after colour error cut. The box shows the colour and
    magnitude region selected for potential globular
clusters. Typical colour and magnitude errors are indicated.}
    \label{f:wcolourselect}
\end{figure}

Various selection criteria were 
applied to the data, i.e. a colour error cut of $\pm$ 0.5 mag., a 
colour selection of $0 < V - I < 2$, and a magnitude selection of 
$20.5 < I < 24$. The colour error is dominated by the uncertainty
in the $V$-band magnitudes which come from a single short F555W image.
The colour selection can be seen in Figure \ref{f:wcolourerror},
and the 
colour-magnitude selection in Figure \ref{f:wcolourselect}. These cuts, as well as a
visual inspection to remove duplicate matches, obvious
galaxies and edge effects 
resulted in a data set of 204 objects.

\section{Keck Telescope Imaging}

\begin{figure}
    \resizebox{0.9\hsize}{!}{\includegraphics{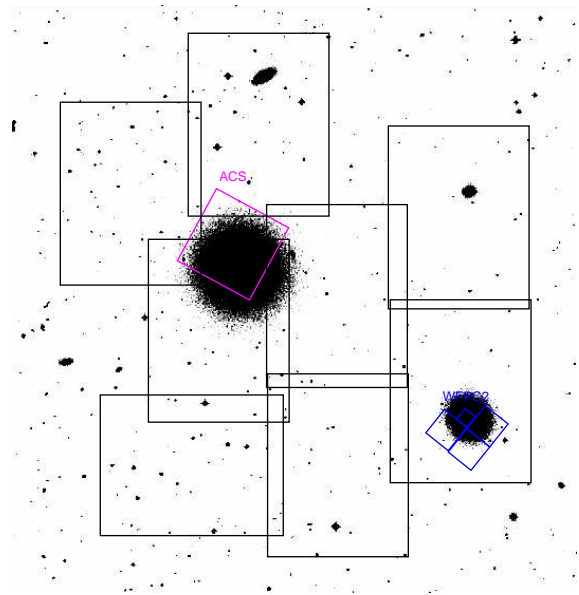}}
    \caption{Position of the eight pointings taken with the Keck
    telescope, the HST/ACS image and the HST/WFPC2 images
    superimposed on a DSS image. North is to the bottom and East is to the left.
	}  
    \label{f:fields}
\end{figure}

\subsection{Observations and Initial Data Reduction}

A mosaic of eight $B$ and $I$-band images of NGC 1407 and the surrounding
area (including NGC 1400)
were acquired using the Low Resolution Imaging Spectrometer
(LRIS; \citealt{b:oke}) on the 10m Keck I telescope 
on 2004 October 12 and 13.
LRIS is equipped with a TEK 2048x2048 detector which is mounted
on the Cassegrain focus with a dual-beam imager so $B$ and $I$ images can 
be obtained simultaneously. 

Eight pointings were obtained
as shown on Figure \ref{f:fields}. Each field
had an area of $\sim$ 6 $\times$  7.8 arcmin$^2$, but since the two CCDs have
slightly different scales (0.138\arcsec pixel$^{-1}$ in $B$ and the
0.215\arcsec pixel$^{-1}$ in $I$), the actual field-of-view
differed slightly between bands. All pointings had an exposure time of
600s. The seeing conditions were not very good and varied between
pointings with a median of $\sim$ 1.3\arcsec.
Time did not allow for a background field to be observed. 
Bias subtraction was performed using the {\sc wmkolris} package provided by the 
observatory and flat-fielding was performed using standard {\sc iraf}
procedures.
  
To aid in the detection of inner globular cluster candidates, a
model of NGC 1407 and NGC 1400 in the $B$-band was subtracted before running our object
detection algorithm (the $I$-band image was saturated). 
A smooth galaxy model was created using
the {\sc ellipse} task in {\sc iraf/stsdas} allowing the centre,
ellipticity and position angle to vary.

\subsection{Object Detection and Selection}
The {\sc daophot} package was used to
select potential globular clusters in each of the eight $B$ and
$I$ 
images. Selection criteria consisted of a roundness range of --1
to 1 and a sharpness range of 0 to 1 with the signal-to-noise
(S/N) ratio threshold, full width half maximum (FWHM) and minimum
and maximum good data values adjusted to the seeing conditions of
each image. This resulted in over 10,000 potential globular
clusters for each filter.

The pointing containing the galaxy centre was adopted as the reference
field to derive photometric zeropoints because a portion of the field 
overlapped with a significant portion of the ACS field. A small
number of objects in common with both the ACS and central Keck
images were identified. The Keck magnitudes were then adjusted to
match the photometrically-calibrated ACS magnitudes, thus giving
zeropoints for the Keck images. 

The {\sc qphot} task in the {\sc apphot} package was used to
measure the magnitude and error of each object found by {\sc daofind}, using an
inner sky annulus radius of the FWHM and a sky annulus width of half the
FWHM and the previously calculated photometric zeropoints.
Spatial matching of the $B$ and $I$ object lists 
reduced the data set to under 5,000 potential GCs. 

\begin{figure}
    \resizebox{0.9\hsize}{!}{\includegraphics[angle=-90]{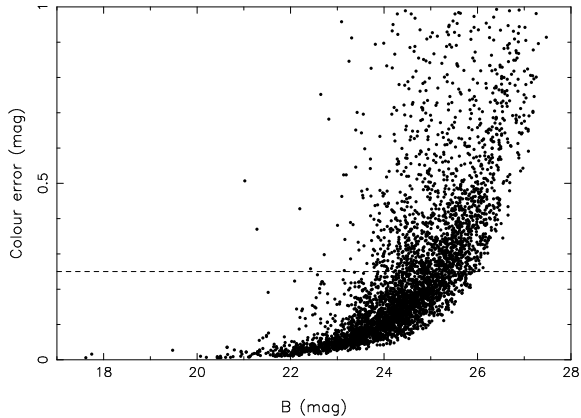}}
    \caption{The colour error as a function of $B$ magnitude for the
    Keck data. The dashed line shows where the selection cut of
$\pm$0.25 mag. was 
    applied.}
    \label{f:kcolourerror}
\end{figure}

\begin{figure}
    \resizebox{0.9\hsize}{!}{\includegraphics[angle=-90]{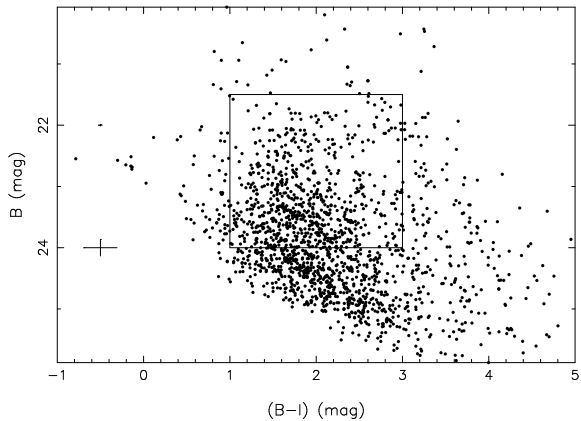}}
    \caption{Colour-magnitude diagram for objects in Keck
    data after roundness and colour error cuts. The box shows the colour and 
    magnitude region selected for potential globular clusters. Typical magnitude and colour errors are shown.}
    \label{f:kcolourselect}
\end{figure}

Various other selection criteria were then applied to the data,
i.e. retaining only objects with roundness between --0.5 and 0.5, colour
error less than $\pm$0.25 mag., $1 < B-I < 3$ and $21.5 < B < 24.0$.
The colour error selection can be seen in Figure
\ref{f:kcolourerror} and the colour and magnitude selection can be seen
in the colour magnitude diagram in Figure \ref{f:kcolourselect}.
These cuts, as well as a visual inspection to remove duplicate matches 
and extended objects resulted in a more manageable and realistic data set of
474 candidate GCs. The bright magnitude cut corresponds to that of
the ACS data, while the colour limits are broader to take into
account the increased photometric error of the Keck data. 

The celestial coordinates of each object were 
calculated using the {\sc Aladin} sky atlas and {\sc Vizier}
catalogue service with the {\sc usno-b1.0} catalogue to identify
four stars in each field. The Starlink {\sc astrom} program was used to
perform the coordinate transformation and fitting, using the
transformation of the central image with the best fit for each
field. 

We determined galactocentric
coordinates for each object based on their celestial coordinates
and the NGC 1407 galaxy centre. We did not detect any GCs within
a galactocentric radius of 20\arcsec.

\subsection{NGC 1400 Globular Clusters}

\begin{figure}
    \resizebox{0.9\hsize}{!}{\includegraphics[angle=-90]{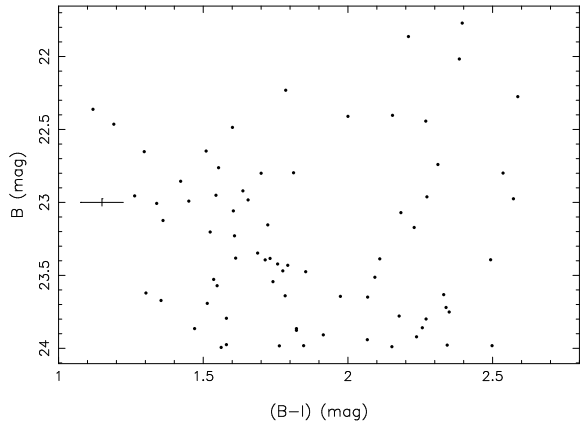}}
    \caption{Colour-magnitude diagram for GC candidates associated 
with NGC 1400 from Keck imaging. 
Typical magnitude and colour errors are shown.}
    \label{f:n1400cmd}
\end{figure}

Visual inspection of the Keck images suggests a population of GCs
associated with NGC 1400. From an analysis of the GC surface
density distribution centred on NGC 1400 (see Section 6.2), we  
associate 74 of the 474 candidate GCs described earlier with NGC 1400.
We have removed these from the NGC 1407 object 
list, leaving 400 candidate GCs associated with NGC 1407. We note that an
extrapolation of the NGC 1407 GC surface density profile (see
Section 5.2 below) would suggest less than 10 GCs associated with
NGC 1407 lie at the projected radius of NGC 1400.
Figure \ref{f:n1400cmd} 
shows a colour magnitude diagram for the 74 GC candidates thought 
to be associated with NGC 1400.

\section{NGC 1407 Results}

\subsection{Globular Cluster Colours} 

\begin{figure}
    \resizebox{0.9\hsize}{!}{\includegraphics[angle=-90]{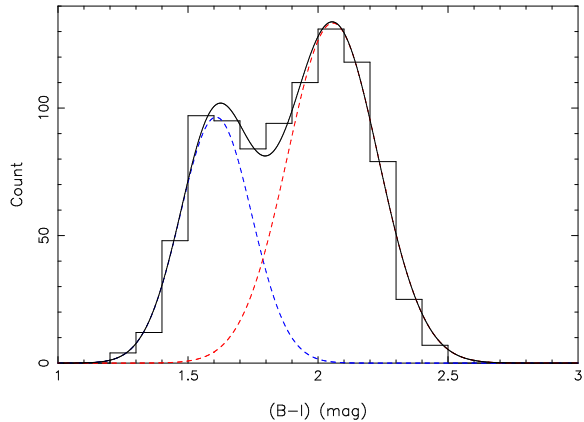}}
    \caption{Colour distribution of NGC 1407 globular clusters
    found in the ACS data. The histogram is binned by 0.1
    magnitudes. The dashed blue and red lines represent a
    Gaussian fit to the blue and red subpopulations respectively,
    with the solid line showing the combined fit. Peaks are
    found at $B-I$ = 1.61 and 2.06}
    \label{f:hgaussfit}
\end{figure}

In Figure \ref{f:hgaussfit}, the colour histogram for the ACS data is
shown, with two Gaussians fitted to the
distribution. This fit was done using the {\sc ngaussfit} task in
{\sc iraf} and allowing the amplitudes and
central position and FHWM to vary. 
The system is well fit by two Gaussians with 36\%
in the blue subpopulation and 64\% in the red one. The blue
subpopulation has a peak colour of $B-I$ = 1.61 and the red
subpopulation peak is at $B-I$ = 2.06; both with an uncertainty 
of $\pm$0.02. 
This is confirmed by a KMM (\citealt{b:ashman}) statistical analysis
on the unbinned colour data which gives peaks of $B-I$ = 1.61 and
2.06, with 38\% in the blue subpopulation and 62\% in the red
one. 

The blue subpopulation, with a Gaussian width $\sigma$ = 0.14 $\pm$
0.03 has a narrower distribution than 
the red subpopulation, with a $\sigma$ = 0.17 $\pm$ 0.03.  
The rms colour error for our sample is $\sim$0.06 which
indicates an intrinsic 
widths of $\sigma$ = 0.13 and
0.16 mag. for the blue and red subpopulations respectively. 
Thus both
subpopulations contain a range of ages and/or metallicities. 
If we assume that the GCs are all very 
old (see Cenarro et al. 2005, in prep.), 
and that the intrinsic colour widths are entirely due to 
metallicity, we can derive the metallicity spread. 
Using the relation of Barmby et al. (2000), we derive 
widths of $\sigma$[Fe/H] = 
0.72 and 0.89 dex for the blue and red subpopulations 
respectively. Any age differences will make these metallicity 
ranges upper limits. ACS imaging of GC systems in other 
galaxies also indicates red subpopulations that are broader 
in colour than the blue subpopulation (Strader et al. 2005; 
Harris et al. 2005). 
We note that equally good fits to the data can be found with
widths that are the same for both subpopulations. In the case of
equal width colour distributions, the relative proportions are
essentially unchanged.

\begin{figure}
    \resizebox{0.9\hsize}{!}{\includegraphics[angle=-90]{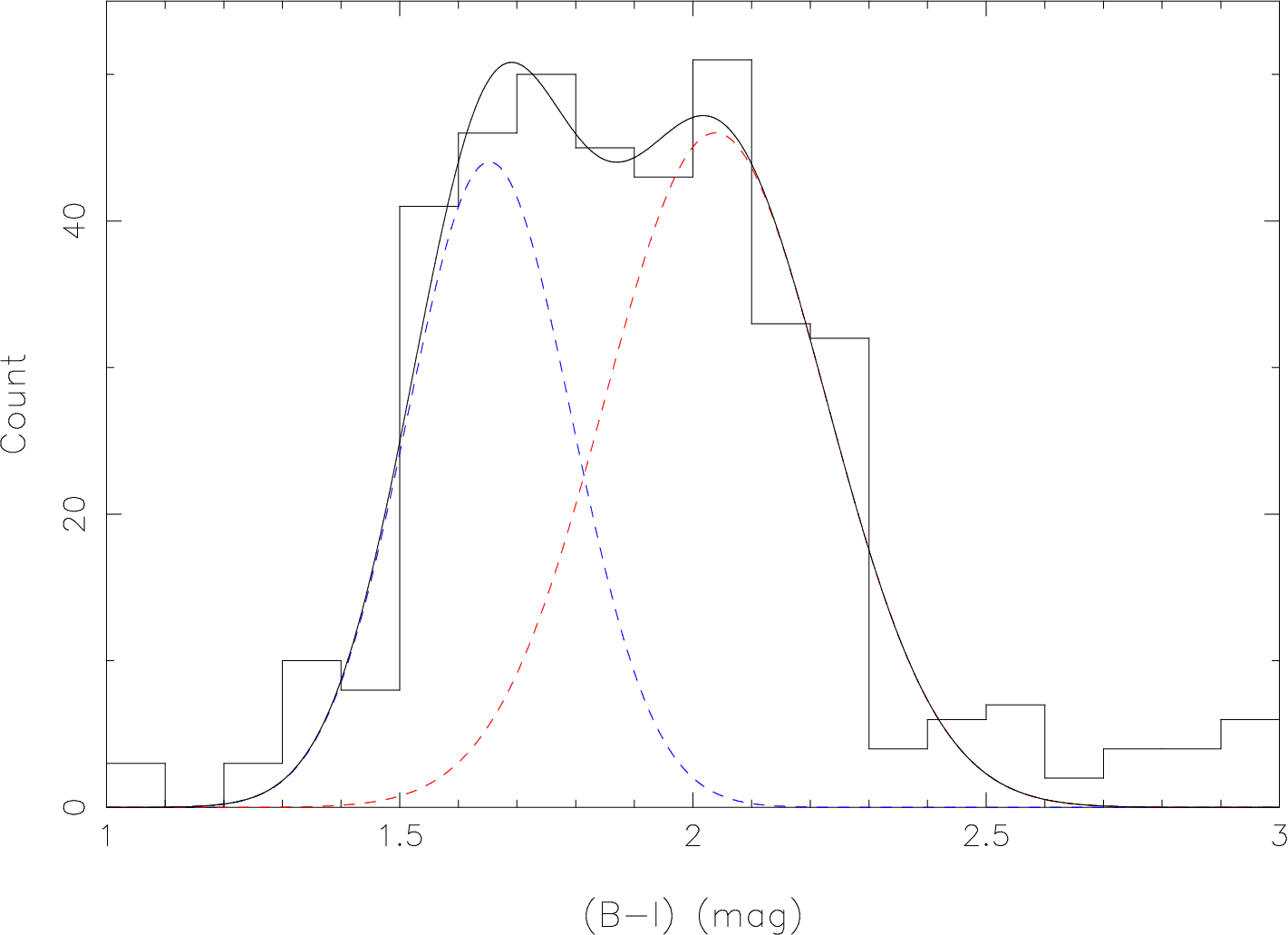}}
    \caption{Colour distribution of NGC 1407 candidate globular clusters
    found in Keck data. The histogram is binned by 0.1
    magnitudes. The dashed blue and red lines represent a
    Gaussian fit to the blue and red subpopulations respectively,
    with the solid line showing the combined fit. Peaks are
    found at $B-I$ = 1.65 and 2.04}
    \label{f:kgaussfit}
\end{figure}

The Keck colour distribution is slightly less well-defined than the ACS 
one (see Fig. \ref{f:kgaussfit}), however we were still able to fit two
Gaussians using the same method. Best-fit Gaussians were found with
peaks at $B-I$ =  1.65 and $B-I$ =  2.04 
$\pm$ 0.05. The blue subpopulation was 41\% and the red one 59\%
of the total, with an uncertainty of about 10\%.  
These peaks and relative proportions are statistically the same
as those found in the ACS data (we might expect to find
relatively more blue GCs in the Keck data as it probes to larger
galactocentric radii). 
Again, we see that the blue subpopulation, with
$\sigma$ = 0.14 is narrower then the red subpopulation, with 
$\sigma$ = 0.19, and that both values are very 
similar to those in the ACS data. We note that the mean colour for the Keck 
data ($B-I$ = 1.92 $\pm$ 0.02) is consistent with that from the ACS data 
($B-I$ = 1.89 $\pm$ 0.01). 

Finally, we note that the bright (B $\le$ 24) blue GCs are on average 
slightly redder (by $\sim$ 0.1 mag.) than the faint blue GCs in
the ACS data. (The red GCs 
do not appear to show a similar colour-magnitude trend.) Although this 
trend is not particularly noticable in our NGC 1407 ACS data, it has been 
seen recently for the GC systems of other galaxies studied with the ACS (Strader 
et al. 2005; Harris et al. 2005). 
The trend is unlikely to be an artifact of photometric errors (see also
Harris et al. 2005) nor is it due to size variations (we find no
strong size-luminosity trend within a GC subpopulation).
Such trends have not been
reported in ground-based or even HST WFPC2 studies of GC systems,
and we find no obvious trend in our Keck data. 

\subsection{Globular Cluster System Surface Density}

\begin{figure}
    \resizebox{0.9\hsize}{!}{\includegraphics[angle=-90]{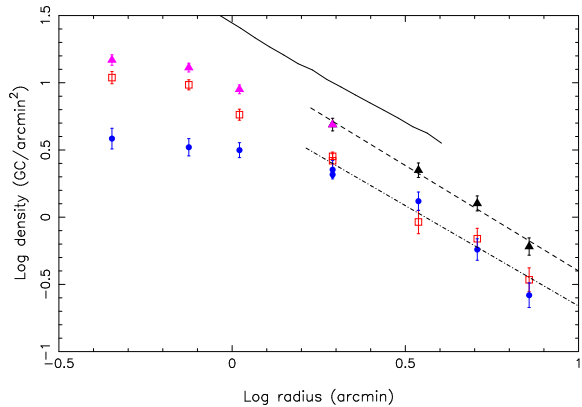}}
    \caption{NGC 1407 globular cluster surface density profiles.
    The total ACS GCs are shown in magenta triangles and the total Keck GCs 
    are shown in black triangles. The red subpopulations are shown in red 
    squares and the blue subpopulations are shown in blue circles. Poisson 
    error bars are given. The Keck data points have been background
    subtracted. The dashed lines show a power-law fit to the
outer region GCs 
and the solid line shows the galaxy surface brightness profile. One arcmin 
is about 6.12 kpc.}
    \label{f:msdensity}
\end{figure}

The surface density versus galactocentric radius of both the ACS 
and Keck data sets (total $\sim$ 1300 objects) is shown in Fig. 
\ref{f:msdensity}. A correction was made for the missing area in
each radial annulus and errors were calculated assuming Poisson
statistics. The outermost ACS annulus was chosen to match the innermost
Keck annulus so that the ACS data could be normalised to
match the Keck data. The subpopulations have been
defined where the two Gaussian fits to their colour distributions
contribute
the same amount to the combined fit, i.e. $B-I \sim$ 1.8 for
both the ACS and Keck data.

The outermost Keck annulus, at a mean radius of 70 kpc is consistent with a 
constant surface density of objects. This suggests that we have   
reached the background and so these values have been 
subtracted from all the other Keck annuli. The ACS data are
assumed to be free from background contamination. 

Each of the data sets (total, blue and red globular cluster populations)
show a centrally concentrated globular cluster system with a
`core' region of near constant surface density and a power-law
like fall-off in the outer regions. Comparing the blue and the red subpopulations, it is noted  
that the blue subpopulation has a much larger core radius than the red 
subpopulation. Following \citet{b:forbes} we
fit the core region with an isothermal-like profile of the form: 
$\rho \approx (r_c^2 + r^2)^{-1}$, where $r_c$ is the core
radius. Fitting only the ACS data points, we 
derive $r_c$ for the total population to be $1.23 \pm 0.09^{'}$,
$0.99 \pm 0.11^{'}$  for the red subpopulation and $2.32 \pm 0.13$
arcmin for the blue subpopulation (where 1 arcmin is 6.12 kpc).

A simple power-law ($\rho \approx r^{-\alpha}$)
fit to the outer most four data points, from a
radius of 1.95\arcmin ~to 7.2\arcmin, gives a slope of --1.57
$\pm$ 0.08 for the total GC system. This can be compared to 
the steeper values found by \citet{b:perrett} of $-2.23 \pm 0.34$ in the 
$T1$-band and $-1.74 \pm 0.38$ in the $I$-band. The slopes for
the blue and red subpopulations are --1.65 $\pm$ 0.29 and --1.50 $\pm$
0.06 respectively, which are statistically the same as each other
and the total GC system. 

Figure \ref{f:msdensity} also shows the galaxy $B$-band surface brightness
profile from the Keck data 
as a solid line, after converting to log units and applying 
an arbitrary vertical
normalisation.  
As can be seen, for the region of
overlap with the outer GC system, the galaxy starlight has a similar 
slope of --1.42 to that of the GC system with slope of --1.57 $\pm$ 0.08.

The plot also shows that there are more red globular clusters in the inner 
parts of the galaxy than
blue.
At larger radii, there appears to be slightly
more red GCs than blue ones, which is an unexpected result as studies
of other galaxies have generally found the blue subpopulation dominating 
the red at large radii. This could be due to increased contamination rates
in the red subpopulation at large radii.

\subsection{Globular Cluster System Position Angle}

\begin{figure}
    \resizebox{0.9\hsize}{!}{\includegraphics[angle=-90]{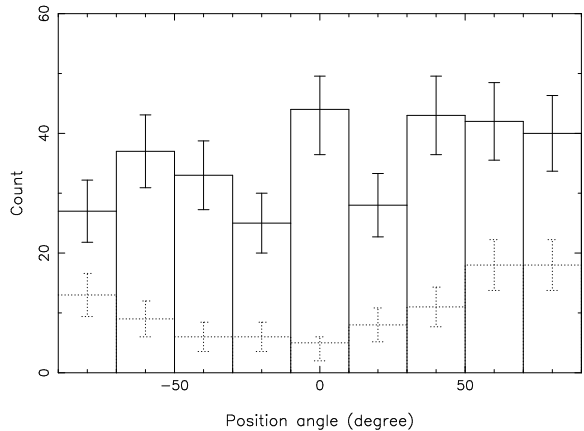}}
    \caption{Histogram of GC position angles for NGC 1407, folded by
halves in bins of 
    twenty degrees.
    Poisson errors are shown. The black solid histogram shows the
ACS data and
    the dotted histogram is the Keck data from a single annulus.}
    \label{f:mangle}
\end{figure}

Figure \ref{f:mangle} shows the position angle dependence of GCs
in NGC 1407. This figure was constructed
from the annuli in which we had complete coverage; for the Keck data this was a
single thin annulus between 2.7\arcmin and 4.2\arcmin ~and for the ACS 
data, it was all GCs within a 0.9\arcmin ~radius of the centre of the 
galaxy. Within the Poisson error bars a flat distribution
(i.e. no dependence on position angle) is consistent with the
data. Since NGC 1407 is an E0 galaxy, it might be expected that the globular cluster
distribution should not depend on position angle. This is indeed
what has been found.

Using the position angle of each globular cluster in the complete coverage
area, we also computed the ellipticity of the GC system using 
$\sqrt{< \langle\sin 2\phi\rangle >^2 + < \langle\cos 2\phi\rangle >^2}$ 
where $\phi$ is the position angle.
We obtained an ellipticity measure of 0.13 $\pm$ 0.08 from the Keck data and 0.06 $\pm$ 0.04
from the ACS data, both corresponding to an E0--E1 distribution. 

\subsection{Globular Cluster Sizes}

\begin{figure}
    \resizebox{0.9\hsize}{!}{\includegraphics[angle=0]{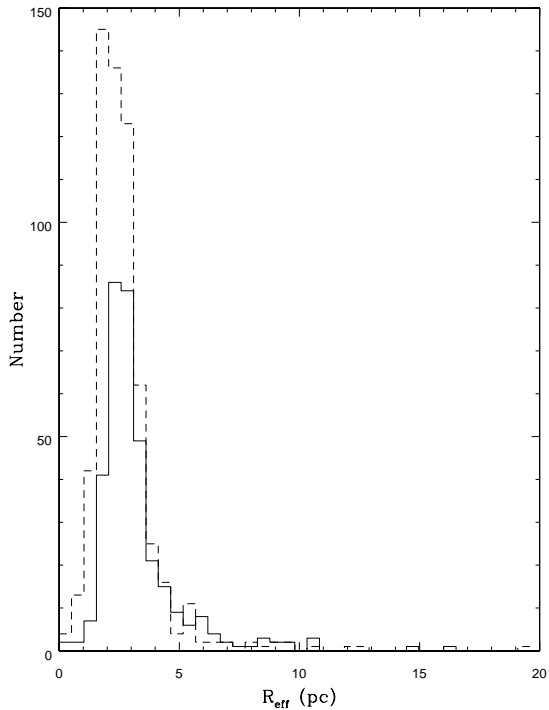}}
    \caption{Histogram of blue and red GC effective radii. The
blue GCs (solid lines) have a mean effective radius of 3.24 $\pm$ 
0.19 pc and the
reds (dashed lines) 2.68 $\pm$ 0.12 pc.} 
    \label{f:size}
\end{figure}

With the superior spatial resolution of the ACS, the GCs in NGC
1407 are marginally resolved. 
We have convolved a Moffat profile with a PSF taken from a star
in the I-band ACS image. Using this PSF with the ISHAPE routine (Larsen
1999), we have determined the size of 
each GC candidate.  
We find a large range in the effective radii of individual GCs 
but the mean value is R$_{eff}$
= 2.9 $\pm$ 0.1 pc. 
This is consistent with the mean value
for the Milky Way GC system of 3.5 $\pm$ 0.4 pc 
(Harris 1996). 
Dividing the sample into blue and red
subpopulations (at B--I = 1.8) we find a statistically
significant difference in their sizes, i.e. for the blue GCs the
mean R$_{eff}$ = 3.24 $\pm$ 0.19 pc vs R$_{eff}$ = 2.68 $\pm$ 0.12 pc
for the red ones. Thus the red GCs appear to be $\sim$20\%
smaller than the blue GCs. A histogram of the blue and red GC
distributions is shown in Figure \ref{f:size}. The distributions
are similar but the red GC distribution is peaked at a smaller
effective radius. 

We also find a half
dozen objects with sizes $\sim$ 20--40 pc (not shown in Figure 
\ref{f:size}). Visual inspection
shows these large objects to be well-detected and similar in appearance 
to other GCs. They cover the full range of magnitudes from B =
22.8 to the magnitude limit. They have a mean colour of B--I = 1.78 $\pm$ 0.11,
i.e. statistically the same as the general GC
population. However, as well as larger sizes they are more
elliptical (less round) than the general GC population.

\subsection{Globular Cluster System Luminosity Function}

The luminosity function of GC systems (number of GCs per unit
magnitude) has been shown to be remarkably similar, or `universal',
across a wide range of galaxies. For example, in a recent review
of GC luminosity functions, Richtler (2003) derived a 
a mean peak, or turnover magnitude in the V-band (which is the
most commonly studied) of M$_V$ = --7.51 $\pm$ 0.24
and $\sigma$ $\sim$ 1.2 for a sample of early-type
galaxies. 

The NGC 1407 GC system has peaks
(V--I)$_o$ $\sim$ 0.95 and 1.15, which implies an I-band turnover
magnitude of M$_I$ = --8.46 and --8.66 for the metal-poor and metal-rich
subpopulations respectively. For the total GC system, we assume
(V--I)$_o$ = 1.05 and hence M$_I$ = --8.56. 
For the B-band, we use (B--V)$_o$ =
0.7 and 0.9 which corresponds to the expected colours for [Fe/H] =
--1.5 and --0.5 of a 15 Gyr old stellar population. This gives M$_B$
= --6.81 and --6.61 for 
the two subpopulations. For the total GC system, we assume
(B--V)$_o$ = 0.8 and hence M$_B$ = --6.71.

We have fit the ACS B and I-band GC luminosity functions to both a
Gaussian and t$_5$ function,  using a maximum likelihood
code developed by J. Secker \citep{b:secker}. This code includes 
our photometric errors and completeness functions (see
Section 2) to make the appropriate correction at faint
magnitudes. The results of 
fitting a Gaussian function are summarised in Table 1 (the
results for a t$_5$ function are within 0.05 of the Gaussian
turnover magnitude for all fits). We have fit the total GC system
as well as the
blue and red subpopulations separately 
(the latter with a fixed $\sigma$ = 1.2).

Our results can be directly compared to those of
\citet{b:perrett} for the I-band. For a fixed
$\sigma$ = 1.2, they found a turnover magnitude of I = 23.12
$\pm$ 0.15 (after adjusting for our I-band extinction
correction). We derive I = 22.98 $\pm$ 0.06  and $\sigma$ = 1.17
$\pm$ 0.05 which is quite consistent with the \citet{b:perrett} findings. 

We find that the B-band turnover magnitude for the metal-rich GCs
is 0.29$^m$ fainter than the metal-poor GCs. However in the
I-band the metal-rich GCs are {\it brighter} than the metal-poor
ones by 0.42$^m$. This could be explained qualitatively if the
metal-rich GCs were on average somewhat younger than the
metal-poor ones (for an old, coeval population the metal-rich GCs
are expected to be fainter in the B-band and be of similar
brightness in the I-band (Ashman, Conti \& Zepf 1995)). We note
that in the V-band the Galactic metal-rich GCs are fainter,
by 0.46 $\pm$ 0.36$^m$,  and for a sample of 13
early-type galaxies they are fainter by 0.30 $\pm$ 0.16$^m$
(Larsen et al. 2001). In terms of the distance modulus, we reassuringly find consistent values
between the B and I-bands.

\begin{table*}
\renewcommand{\arraystretch}{1.0}
\begin{tabular}{lccc}
\multicolumn{4}{c}{{\bf Table 1.} NGC~1407 Globular Cluster
Luminosity Function}\\
\hline
Total & M$_B$ = --6.71 & M$_I$ = --8.56 &\\
\hline
B = 24.95 $\pm$ 0.08 & m--M = 31.63 $\pm$ 0.09 & $\sigma$ = 1.21 $\pm$ 0.05
& N$_{GC}$ = 1171\\
I = 22.98 $\pm$ 0.06 & m--M = 31.51 $\pm$ 0.07 & $\sigma$ = 1.17 $\pm$ 0.05
& N$_{GC}$ = 1148\\
\hline
Blue & M$_B$ = --6.81 & M$_I$ = --8.46 &\\
\hline
B = 24.76 $\pm$ 0.09 & m--M = 31.57 $\pm$ 0.10 & $\sigma$ = 1.2 
& N$_{GC}$ = 419\\
I = 23.29 $\pm$ 0.09 & m--M = 31.75 $\pm$ 0.10 & $\sigma$ = 1.2 
& N$_{GC}$ = 473\\
\hline
Red & M$_B$ = --6.61 & M$_I$ = --8.66 &\\
\hline
B = 25.05 $\pm$ 0.08 & m--M = 31.66 $\pm$ 0.09 & $\sigma$ = 1.2
& N$_{GC}$ = 753\\
I = 22.87 $\pm$ 0.06 & m--M = 31.53 $\pm$ 0.07 & $\sigma$ = 1.2
& N$_{GC}$ = 696\\
\hline
\end{tabular}
\\Notes: The table gives the turnover magnitude, 
distance modulus, Gaussian dispersion and total number of GCs for
the total, blue (metal-poor) and red (metal-rich)
populations. The assumed absolute magnitudes are quoted for the B
and I bands. For the blue and red fits, the Gaussian width has
been fixed to 1.2.  
\end{table*}

When we examine the the total, blue and
red (divided at B--I = 1.8) 
populations separately we measure 31.57, 31.66 and 31.60 respectively. The uncertainty on these
three estimates are similar at about $\pm$ 0.1$^m$.  We take as
our best distance modulus value 31.6 $\pm$ 0.1. 

\citet{b:richtler} recently concluded that GC luminosity
functions ``...are as accurate as those derived from surface
brightness fluctuations, once the conditions of high data quality
and sufficiently rich cluster systems are fulfilled.''
Interestingly the surface brightness fluctuation
(SBF) distance for NGC 1407 appears to be poorly constrained. 
The SBF distance modulus for NGC 1407 was given in Tonry et al. (1991) as
31.07 $\pm$ 0.13. This was revised to 32.30 $\pm$ 0.26 in Tonry
et al. (2001). Jensen et al. (2003) recalibrated the Tonry et
al. (2001) SBF values by --0.16, thus giving 32.14 $\pm$ 0.13. 
A recent SBF analysis of the same ACS
data used here (Cantiello et al. 2005) suggests a value of 32.00 $\pm$ 0.1. 

We note that the
D$_n$--$\sigma$ velocity to NGC 1407 is 1990 $\pm$ 187 km/s (Faber et
al. 1989), which for H$_o$ = 72 km/s/Mpc implies a distance of
27.6 $\pm$ 2.6 Mpc (m--M = 32.2 $\pm$ 0.2). 
This is consistent with the larger SBF value. The distance
modulus from the Faber-Jackson relation from HyperLeda (i.e. m--M = B$_T$ + 6.2
log$\sigma$ + 5.9) is 31.59. Using 
the Virgo infall corrected velocity for NGC 1407 of 1617 km/s,
and a Hubble constant of H$_o$ = 72 km/s/Mpc, gives 
22.5 Mpc or m--M = 31.76. The latter two measures are 
similar to the distance from the GC
luminosity function. 
As far as we aware, a distance estimate based on the planetary
nebulae luminosity function does not yet exist for NGC 1407. We
conclude that NGC 1407 lies at a distance of $\sim$31.6 rather
than $\sim$32.1. 
If we had assumed the larger distance of 26.3 Mpc, then all of
the galactocentric radii and GC sizes quoted in this paper need
to be corrected by a factor of 25\%. 

After examining the value of the turnover magnitude in several radial bins, we 
find no evidence for a turnover magnitude that varies with
galactocentric radius. Integrating under the GC luminosity function gives the 
total number of GCs, within the ACS field-of-view, to be about
1160. Using photometry from NED we estimate a V-band
magnitude within the ACS field-of-view to be V $\sim$ 10.3. This
gives a GC specific frequency of S$_N$ = 3.5, which as a lower
limit to the total S$_N$ is consistent with the value of 4.0
$\pm$ 1.3 found by Perrett et al. (1997).

\section{NGC 1400 Results}

\subsection{Globular Cluster Colours}

\begin{figure}
    \resizebox{0.9\hsize}{!}{\includegraphics[angle=-90]{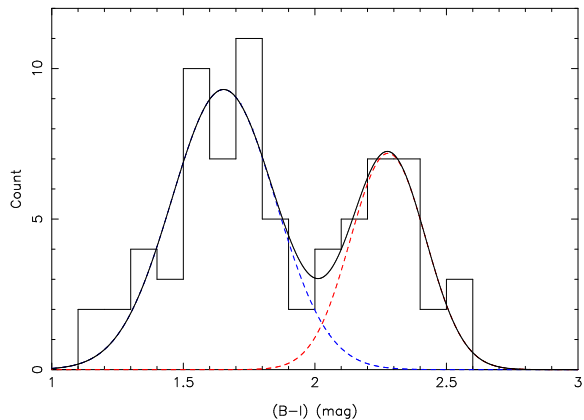}}
    \caption{NGC 1400 GC colour distribution from Keck data. The
histogram is binned by 0.1 magnitudes. The dashed
    blue and red lines represent a Gaussian fit to the blue and red
    subpopulations respectively, with the solid line showing the combined fit.
    Peaks are found at $B - I$ = 1.65 and 2.28.}
    \label{f:cgaussfit}
\end{figure}

We only associated 74 GCs with NGC 1400 from the Keck images, so any subsequent
analysis will be subject to small number statistics.
A GC colour distribution is shown in Fig. \ref{f:cgaussfit} with two 
Gaussians fitted to the distribution. The blue subpopulation has a peak colour
of $B - I$ = 1.65 and the red subpopulation has a peak colour of $B - I$ =
2.28. The uncertainty in estimating these peak values is 
$\pm$ 0.1 mag. The blue subpopulation contains 71\% of the total,
and the red 29\%.
A KMM analysis supports this finding with peaks at
$B-I$ =  1.62 and 2.29 in the ratio of 64\% blue to 36\% red. 
The blue peak is similar to that found in NGC 1407
but the red peak seems substantially redder. 
Small number statistics and contamination (particularly in the
red subpopulation) can
strongly affect the mean colours. Nevertheless, bimodality in the NGC
1400 GC colours seems reliable. 

For peaks of $B-I$ = 1.65 and 2.28, coupled
with the conversion formula of \citet{b:forte} we would predict
$V-I$ peaks at 0.95 and 1.27 (a red peak at $B-I$ = 2.06 would
give $V-I$ = 1.16). 
In Fig. \ref{f:wcolourhistm} we show the $V-I$ colour distribution
for the $\sim$ 200 GCs detected in the WFPC2 images. The
distribution shows a peak around $V-I \sim$ 1.1.
A KMM analysis does
not detect two peaks in the distribution, suggesting a unimodal
colour distribution. 
The metallicity sensitivity of V--I is much less than
B--I, so it is possible that the two distributions are consistent
with each other. To test this we have performed 
the exercise of converting our Keck B--I distribution into V--I and convolving 
with the appropriate V--I colour errors. This model distribution
is also shown in Fig. \ref{f:wcolourhistm} and is 
consistent with the observed V--I distribution according to a KS
test. Thus the use of V--I colours can obscure the presence of a
bimodal colour distribution (as seen in B--I).

\begin{figure}
    \resizebox{0.9\hsize}{!}{\includegraphics[angle=-90]{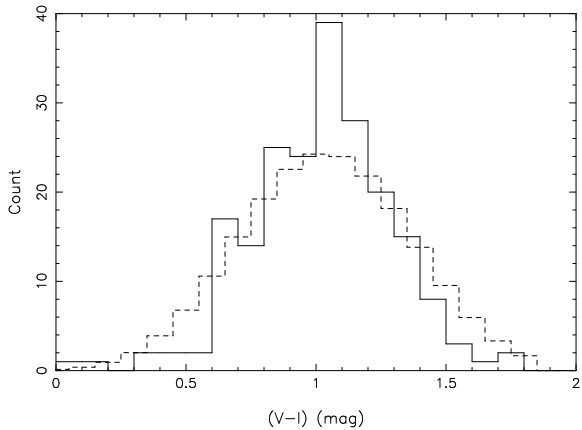}}
    \caption{Colour distribution of NGC 1400 globular clusters
    found in the HST/WFPC2 data. The histogram is binned by 0.1
    magnitudes. The solid histogram shows the observed V--I 
distribution, and the dashed histogram shows a model based on the
convolved B--I colours (see text for details). 
}
    \label{f:wcolourhist}
\end{figure}

\subsection{Globular Cluster System Surface Density}

\begin{figure}
    \resizebox{0.9\hsize}{!}{\includegraphics[angle=-90]{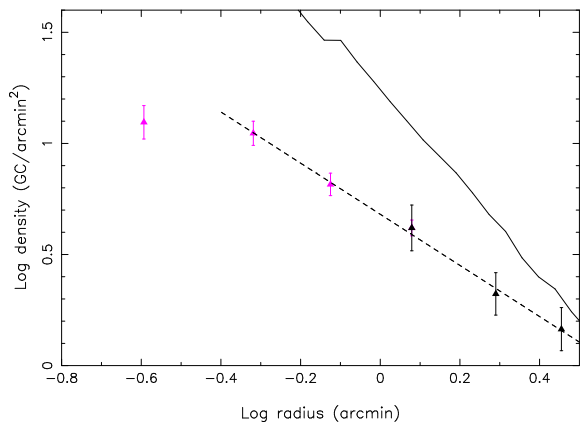}}
    \caption{NGC 1400 surface density profile with Poisson error
bars. The Keck data have been 
    background subtracted, and the WFPC2 data normalised to the
innermost Keck data point. The dashed line shows a power-law fit to the data 
    with a slope of --1.15 and the solid line shows the galaxy surface
    brightness profile. One arcmin is about 6.12 kpc.}
    \label{f:csdensity}
\end{figure}

In Figure \ref{f:csdensity} we show a plot of surface density versus 
galactocentric radius for NGC 1400 GCs. The local background density calculated 
previously has been subtracted from the Keck data points (the
WFPC2 data points are assumed to have negligible background contamination). A correction has been made for
the missing area in each radial annulus. The profile shows a near
constant core region and rapid fall-off in the outer parts,
similar to that seen in NGC 1407. An isothermal core profile fit
to the WFPC2 data points gives a core radius $r_c$ = 0.76 $\pm$ 0.10
arcmin (4.7 $\pm$ 0.8 kpc).  
A simple power-law fit ($\rho \approx r^\alpha$), excluding the
inner most point, gives a slope of --1.15 $\pm$
0.01. \citet{b:perrett} found  --2.35 $\pm$ 0.78 in the $T1$-band
and --1.57 $\pm$ 0.34 in the $I$-band. 
The declining slope suggests that the NGC 1400 GC system extends
out to at least a  galactocentric radius of 3.3\arcmin ~(20 kpc). 
The figure also shows the
NGC 1400 $B$-band surface brightness profile from the Keck data  
converted to logarithm space with an arbitrary vertical normalisation applied.
The galaxy starlight gradient is --1.88, which is steeper than that of the GCs.

\subsection{Globular Cluster System Position Angle}

\begin{figure}
    \resizebox{0.9\hsize}{!}{\includegraphics[angle=-90]{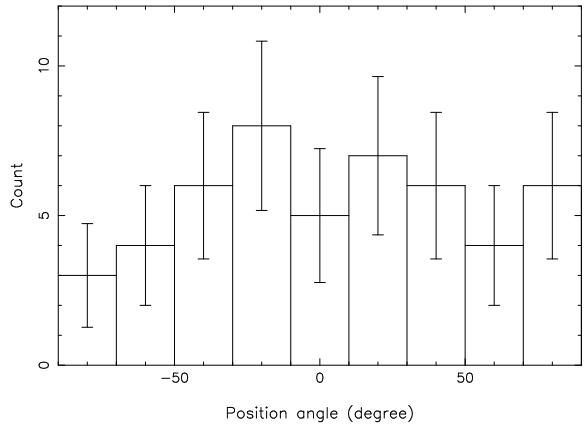}}
    \caption{
Histogram of GC position angles for NGC 1400, folded by
halves in bins of 
    twenty degrees 
    with Poisson error bars.}
    \label{f:cangle}
\end{figure}

A histogram of folded position angle from the Keck data is given in Figure \ref{f:cangle} which
shows a fairly flat distribution, within the errors. This was constructed using objects within a
2.6\arcmin ~radius of the centre of the galaxy, for which we had complete
coverage. Thus like NGC 1407, the GC distribution around NGC 1400
is consistent with no preferred position angle as might be
expected for a nearly circular host galaxy. This is further reinforced by an
ellipticity calculation which yielded 0.2 $\pm$ 0.1, corresponding to an E2
distribution.

\section{Discussion and Conclusions}

From our combined Keck and Hubble Space Telescope imaging we have
measured several properties of the globular cluster systems
associated with NGC 1407 and NGC 1400 in the Eridanus cloud. We
have detected bimodality in the GC colour distribution for both
galaxies. This indicates the presence of two
GC subpopulations in these galaxies, similar to the situation for
most well studied early-type galaxies. 

For NGC 1407, we clearly detect
bimodality with a blue (metal-poor)
40\% subpopulation of mean colour of $B-I$ = 1.61 $\pm$ 0.02, and a
red (metal-rich) 60\% subpopulation with $B-I$ = 2.06 $\pm$
0.02 from HST/ACS imaging. Using the relation of Barmby et
al. (2000) to transform colour into metallicity (i.e. assuming
that the NGC 1407 GCs are predominately old), then the
subpopulations correspond to [Fe/H] = --1.45 and --0.19. 
The Keck imaging results are less
obviously bimodal but consistent with the ACS results.

For NGC 1400 we detect bimodality in the Keck data at $B-I$
$\sim$ 1.65 and 2.28, however with only 74 candidate GCs detected
such values could be subject to systematic offset. In the
HST/WFPC2 dataset we have over 200 GCs, however the $V-I$ colour
is much less sensitive to metallicity than $B-I$ making it more
difficult to detect any bimodality. The colour distribution is
formally unimodal but hints at peaks around $V-I \sim$ 0.9 and
1.1 $\pm$ 0.05. This would correspond to $B-I$ = 1.55 $\pm$ 0.1
and 1.95 $\pm$ 0.1.

Such colour peaks are reminiscent of the GC systems of NGC 1399 
and NGC 1404 in the Fornax cluster \citep{b:grillmair}. 
These galaxies have similar luminosities (M$_V$ = --22.00 and
M$_V$ = --21.46 respectively for m--M = 31.49) as NGC 1407 and
NGC 1400. GC colour peaks were found at $B-I$ = 1.7 and 2.1 $\pm$
0.05 for NGC 1399 and 1.6 and 2.1 $\pm$ 0.05 for NGC 1404,
i.e. similar to those seen in NGC 1407 and NGC 1400.

We can compare the subpopulation mean colours with predictions
from the colour-galaxy luminosity scaling relation of \citet{b:strader} after
conversion $V-I$. For NGC 1407,  peaks at $B-I$ = 1.61
and 2.06 correspond to $V-I$ = 0.93 and 1.16 using the transformation of 
Forbes \& Forte (2001). These values are very consistent with that expected 
for a M$_V$ = --21.86 galaxy. For NGC 1400, the blue peak at $B-I$ = 1.62 or 
$V-I$ = 0.94 is also consistent for its luminosity of M$_V$ = --20.63. 
However, the red peak at around $B-I$ = 2.29 or $V-I$ = 1.28 is much redder 
than the \citet{b:strader} relation. This suggests that the red subpopulation 
in NGC 1400 is poorly defined in terms of its peak colour.  

In addition to measuring the mean colour of the two
subpopulations, we have also measured the width of the subpopulations
for NGC 1407. We find that both subpopulations 
are intrinsically broad (indicating a range of ages and/or
metallicities), with the red subpopulation being broader than the blue.  

We find the half--light radius of red GCs to be on average 20\%
smaller than those of blue GCs. 
Smaller relative sizes for the red GCs has been reported for several
galaxies (Kundu \& Whitmore 1998; Larsen et al. 2001).  
Larsen \& Brodie (2003) have speculated that the sizes differences may
not be real but due to projection effects. If so, they predict a
relatively strong size--galactocentric radius relation. We find no
evidence for such a trend in our data, in either the combined or
blue and red GC subpopulations separately. Although we note that 
a careful modeling of the projected and 3D relations is needed to
conclusively rule out this possibility. 
An alternative
explanation, that predicts no size--galactocentric radius trend
is that proposed by Jordan (2004). He suggests that the process
of mass segregation and stellar evolution effects will result in
a smaller measured half--light radius for the metal-rich
GCs. Assuming coeval populations, an initial mass function, 
Michie-King isotropic models
and constant {\it half--mass} radii, his favoured model predicts that the
red subpopulation of NGC 1407 ([Fe/H] = --0.19) will have half--light
radii that are 20\% smaller than the blue subpopulation ([Fe/H] =
--1.45). This is entirely consistent with our measurements. Jordan
notes that if the metal-rich GCs were younger by 3 Gyrs, then the
predicted size difference would increase to 30\%. Thus within the
assumptions of the model, our size measurements suggest that the
GC system of NGC 1407 is old and coeval.

We also find half a dozen objects with effective radii of
20--40 pc. They may be related to
the `faint fuzzies' seen in some S0 galaxies (Larsen \& Brodie
2000). However the faint fuzzies tend to be quite red, whereas
these large objects in NGC 1407 are not particularly red. 
Alternatively, they could be related to Ultra Compact Dwarfs (UCDs)
which have been observed around some elliptical galaxies
(e.g. Drinkwater et al. 2004; Richtler 2005). Although our large objects do not
have the bright magnitudes expected of UCDs associated with NGC
1407. Given their large sizes and elongations, we suspect they
are background galaxies.

The GC surface density profiles of both galaxies reveal a near
constant density central region with a power-law like fall off
in the outer parts, similar to those seen in other GC systems.
We have fit the central region with an isothermal `core' profile
deriving a GC core radius of 7.5 $\pm$ 0.7 kpc for NGC 1407 and
4.6 $\pm$ 0.8 kpc for NGC 1400. These values can be compared to
the GC core radius-galaxy luminosity scaling relation found by
\citet{b:forbes}. Both NGC 1407 and NGC 1400 would appear to have
larger core radii than galaxies of similar luminosity. 
Such core radii in GC systems may be due to destruction via 
tidal shocks \citep{b:vesperini}. 
However, we find no evidence for a mean GC magnitude 
that varies with galactocentric radius as might be
expected. An alternative explanation for the GC system core
region is that of dissipationless merger
events or accretions \citep{b:bekki}. 
 
For NGC 1407, we had sufficient GC numbers to separate the blue
and red subpopulations. We found the red subpopulation to have a
core radius of 6.1 $\pm$ 0.8 kpc and outer slope of --1.50 $\pm$
0.06, while the blue one was somewhat larger
with 14.2 $\pm$ 1.0 kpc and slope of --1.65 $\pm$ 0.29. 
This difference in subpopulation core size, if found
in a number of galaxies, needs to be understood. 

Beyond the central core region the GC surface density slopes were
compared to the galaxy starlight. For NGC 1407 the GC and galaxy
starlight slopes were very similar, around --1.5. However, the GC profile
for NGC 1400 (from the limited Keck data) appears to be 
significantly flatter than the galaxy slope. 
Both galaxies reveal GC slopes that lie within the cosmic scatter
of the GC slope-galaxy luminosity scaling relation \citep{b:harris6}.

For both NGC 1407 and NGC 1400 we find that their GC systems have
a similar ellipticity (i.e. near zero) and azimuthal distribution
(i.e. no strong position angle dependence) as the underlying
galaxy starlight. Similar trends have been seen in other galaxies
(e.g. \citealt{b:forbes}). 

Using the ACS data for NGC 1407, we have fit the GC luminosity
function taking into account photometric errors and
incompleteness. Our I-band fits agree very well with those of
\citet{b:perrett}. As well as fitting the total GC system, we fit
the blue and red subpopulations separately. In the B-band, we find
the red GC subpopulation to have have a turnover magnitude that
is 0.29 magnitudes fainter than the blue one. In the I-band the
turnover magnitudes are similar. 

After applying 
appropriate colour corrections, the accuracy of the
distance modulus is not significantly changed using the total GC
system or one of the subpopulations. 
We derive a distance modulus from the GC luminosity function of
31.6 $\pm$ 0.1. This is similar to distance estimates from the
Faber-Jackson relation and the Hubble distance for H$_o$ = 72
km/s/Mpc. However, this distance lies at the midpoint of recent SBF
distance determinations (i.e m--M = 31.0 and 32.2) and is
inconsistent with the D$_n$--$\sigma$ distance of 32.2 $\pm$ 0.2.

\section{Acknowledgments}

We thank M. Beasley for his help obtaining the Keck images, and
S. Larsen for useful comments on the text. Both
DF and PSB thank the ARC for financial support. JB, JS and LS
thank the NSF grant AST 0206139 for financial support. Finally we thank 
the referee, W. Harris, for several useful suggestions that have improved 
the paper.

\onecolumn

\appendix
\section{Globular Cluster Lists}

\begin{center}
\begin{longtable}{cccccccc}
\caption
[Globular clusters in the HST/ACS field of NGC 1407]
{Globular clusters in the HST/ACS field of NGC 1407} \\
\hline
ID & R.A. & Dec. & $R_{\rm GC}$ & $I$ & $I$ err & $B-I$ & $B-I$ err \\
& (J2000) & (J2000) & (arcsec) & (mag) & (mag) & (mag) & (mag) \\
\hline
\endfirsthead
\multicolumn{8}{c}
{\tablename\ \thetable{} -- continued from previous page} \\
\hline
ID & R.A. & Dec. & $R_{\rm GC}$ & $I$ & $I$ err & $B-I$ & $B-I$ err \\
& (J2000) & (J2000) & (arcmin) & (mag) & (mag) & (mag) & (mag) \\
\hline
\endhead
\hline
\multicolumn{8}{r}{{Continued on next page}} \\
\hline
\endfoot
\hline \hline
\endlastfoot
\hline
\end{longtable}

\begin{longtable}{cccccccc}
\caption
[Globular cluster candidates in the Keck field of NGC 1407]
{Globular cluster candidates in the Keck field of NGC 1407} \\
\hline
ID & R.A. & Dec. & $R_{\rm GC}$ & $I$ & $I$ err & $B-I$ & $B-I$ err \\
& (J2000) & (J2000) & (arcmin) & (mag) & (mag) & (mag) & (mag) \\
\hline
\endfirsthead
\multicolumn{8}{c}
{\tablename\ \thetable{} -- continued from previous page} \\
\hline
ID & R.A. & Dec. & $R_{\rm GC}$ & $I$ & $I$ err & $B-I$ & $B-I$ err \\
& (J2000) & (J2000) & (arcmin) & (mag) & (mag) & (mag) & (mag) \\
\hline
\endhead
\hline
\multicolumn{8}{r}{{Continued on next page}} \\
\hline
\endfoot
\hline \hline
\endlastfoot
\hline
\end{longtable}

\begin{longtable}{cccccccc}
\caption
[Globular clusters in the HST/WFPC2 field of NGC 1400]
{Globular clusters in the HST/WFPC2 field of NGC 1400} \\
\hline
ID & R.A. & Dec. & $R_{\rm GC}$ & $I$ & $I$ err & $V-I$ & $V-I$ err \\
& (J2000) & (J2000) & (arcsec) & (mag) & (mag) & (mag) & (mag) \\
\hline
\endfirsthead
\multicolumn{8}{c}
{\tablename\ \thetable{} -- continued from previous page} \\
\hline
ID & R.A. & Dec. & $R_{\rm GC}$ & $I$ & $I$ err & $V-I$ & $V-I$ err \\
& (J2000) & (J2000) & (arcsec) & (mag) & (mag) & (mag) & (mag) \\
\hline
\endhead
\hline
\multicolumn{8}{r}{{Continued on next page}} \\
\hline
\endfoot
\hline \hline
\endlastfoot
\hline
\end{longtable}

\begin{longtable}{cccccccc}
\caption
[Globular cluster candidates in the Keck field of NGC 1400]
{Globular cluster candidates in the Keck field of NGC 1400} \\
\hline
ID & R.A. & Dec. & $R_{\rm GC}$ & $I$ & $I$ err & $B-I$ & $B-I$ err \\
& (J2000) & (J2000) & (arcmin) & (mag) & (mag) & (mag) & (mag) \\
\hline
\endfirsthead
\multicolumn{8}{c}
{\tablename\ \thetable{} -- continued from previous page} \\
\hline
ID & R.A. & Dec. & $R_{\rm GC}$ & $I$ & $I$ err & $B-I$ & $B-I$ err \\
& (J2000) & (J2000) & (arcmin) & (mag) & (mag) & (mag) & (mag) \\
\hline
\endhead
\hline
\multicolumn{8}{r}{{Continued on next page}} \\
\hline
\endfoot
\hline \hline
\endlastfoot
\hline
\end{longtable}

\end{center}

\end{document}